\documentstyle[prb,epsfig,aps]{revtex}

\begin{document}

\twocolumn[

\hsize\textwidth\columnwidth\hsize\csname @twocolumnfalse\endcsname

\draft
\preprint{\today}
\title{
Far-infrared study of  the two-dimensional 
dimer  spin system SrCu$_2$(BO$_3$)$_2$
}

\author{T. R\~o\~om,  U.~Nagel, E.~Lippmaa}

\address{National Institute of Chemical Physics and Biophysics, Akadeemia
	tee 23, 12618 Tallinn, Estonia}

\author{H.~Kageyama, K.~Onizuka, Y.~Ueda}

\address{Institute for Solid State Physics, University of Tokyo,
	Roppongi, Minato-ku, Tokyo 106-8666, Japan}

\maketitle

\begin{abstract}

Using far-infrared spectroscopy in magnetic fields up to 12\,T we have
studied a two-dimensional dimer spin  gap system SrCu$_2$(BO$_3$)$_2$.
We found several  infrared active modes in the dimerized state (below 10\,K)
in the frequency range from 3 to 100\,cm$^{-1}$.
The measured  splitting from the ground state to the  excited triplet
$M_S=0$ sublevel
is  $\Delta_1=24.2$\,cm$^{-1}$
and the other two triplet state sublevels in zero magnetic field  are  1.4\,cm$^{-1}$ below and
above the $M_S=0$ sublevel.
Another  multiplet is at $\Delta_2=37.6$\,cm$^{-1}$ from the ground state.
A strong    electric dipole  active transition polarized in the 
(ab)-plane  is activated in the dimer spin system below 15\,K  at 52\,cm$^{-1}$.

\end{abstract}


]

\section{Introduction}

The discovery of high temperature superconductivity  in lightly doped
antiferromagnets has
renewed interest in low dimensional spin systems with specific spatial
structures.
Recently a new two-dimensional
spin gap system  SrCu$_2$(BO$_3$)$_2$ was found\cite{Kageyama99}.
The SrCu$_2$(BO$_3$)$_2$  is interesting {\em{first}}, because the ground
state
is known.
SrCu$_2$(BO$_3$)$_2$ has a tetragonal structure  where    structural dimers
of  Cu$^{2+}$ ions
form an orthogonal network
in the CuBO$_3$ planes, separated by Sr atoms\cite{Smith91}.
It is topologically equivalent to a system where the singlet dimer state
is an exact eigenstate of the spin Hamiltonian\cite{Shastry81,Miyahara99}.
{\em{Second}}, for the first time among two-dimensional spin systems
the quantized magnetization plateaus were observed\cite{Kageyama99}.
{\em{Third}}, SrCu$_2$(BO$_3$)$_2$ is close to the quantum critical
transition point,
$J'/J=0.7,$
from the gapful magnetic dimer state to the antiferromagnetically ordered
gapless  state\cite{Miyahara99,Weihong99,Hartmann2000}.
Fit of magnetic susceptibility data gives $J=100$\,K and $J^{'}=68$\,K for
the nearest neighbor
and next-nearest neighbor antiferromagnetic coupling constants between
copper
$S=1/2$ spins.

Despite the well-defined   ground state, little is known about the excited
states.
The distance from the ground singlet state to the  excited triplet state
estimated from  copper nuclear magnetic relaxation rate in a powdered sample
is $\Delta_R=30$\,K.
Submillimeter wave electron spin resonance  (ESR) study\cite{Nojiri99}  on
monocrystals  shows that the
excited triplet state  splits
into two triplet modes and the averaged value for the spin gap is 34.7\,K.
According to the same ESR study  there must be another spin  excitation
above the triplet mode.

Infrared spectroscopy is a good method for studying excitations in a
magnetic system because it
is sensitive to spin, charge and lattice  degrees of freedom.
We present here the results of a low temperature (4 to 45\,K) far-infrared study
of SrCu$_2$(BO$_3$)$_2$ in magnetic fields
from 0 to 12\,T and at frequencies in the 3 to  100\,cm$^{-1}$ range.
We have determined the zero field splitting of the  excited triplet state
and have found  a new strong 
infrared active resonance at 52\,cm$^{-1}$.

\begin{figure}
\begin{center}\mbox{ \epsfig{file=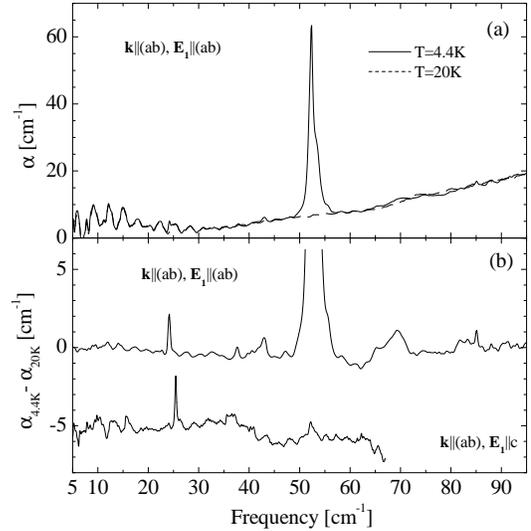, width=7cm, 
	clip=} }\end{center}

\caption{
Infrared absorption spectra of SrCu$_2$(BO$_3$)$_2$
for different light  polarizations.
(a) -- absolute spectra in dimerized state (4.4\,K) and in para\-magnetic
state (20\,K).
(b) -- difference of 4.4\,K and 20\,K spectra.
Spectrum in $\mathbf{k}\!\parallel\!\mathrm{(ab)}$, $\mathbf{E}_1\!\parallel
\!\mathbf{c}$ geometry
has been lowered in vertical direction by $5$\,cm$^{-1}$.
}
\end{figure}

We studied two single crystals of  SrCu$_2$(BO$_3$)$_2$.
The first sample was 0.17mm thick in
the c-direction and had an area of 10\,mm$^2$ in (ab)-plane
(CuBO$_3$ plane).
This sample was used in the measurements where the light
$\mathbf{k}$ vector   was perpendicular to
the (ab)-plane and electric field $\mathbf{E_1}$ polarized in the
(ab)-plane.
The second sample consisted of two pieces 0.65mm thick in a-direction     and
with a total area of
11.5\,mm$^2$ in the (ac)-plane.  This sample was used for measurements where
the light \textbf{k} vector was
in the (ab)-plane and \textbf{E$_1$} either parallel to the c-axis or to
the (ab)-plane.

Far-infrared measurements were done with a polarizing
Martin-Pupplett Fourier transform spectrometer\cite{Sciencetech}
and a  sample cryostat equipped with a 12\,T Oxford Instruments magnet and
two
$^3$He cooled  silicon bolometers from Infrared Laboratories.
Spectra were recorded at 0.4\,cm$^{-1}$ resolution.
The magnetic  field was applied parallel to the direction of light propagation.
Absolute absorption spectra $\alpha(\omega,T)$ were calculated from the transmission 
taking into account two  back reflections
from  sample to vacuum interface: $\alpha(\omega,T)=-d^{-1}\ln[I(\omega,T)/I_0(\omega)(1-R)^2]$,
where $I_0(\omega)$ is the intensity of the incident and $I(\omega,T)$ the intensity of the transmitted
infrared radiation at frequency $\omega$; $d$ is the thickness of the crystal. 
The reflection coefficient $R=0.33$
was calculated from the refraction index $n=3.7$ determined from the fringe
pattern of the 0.17\,mm thick sample transmission spectrum.

\begin{figure}
\begin{center}\mbox{ \epsfig{file=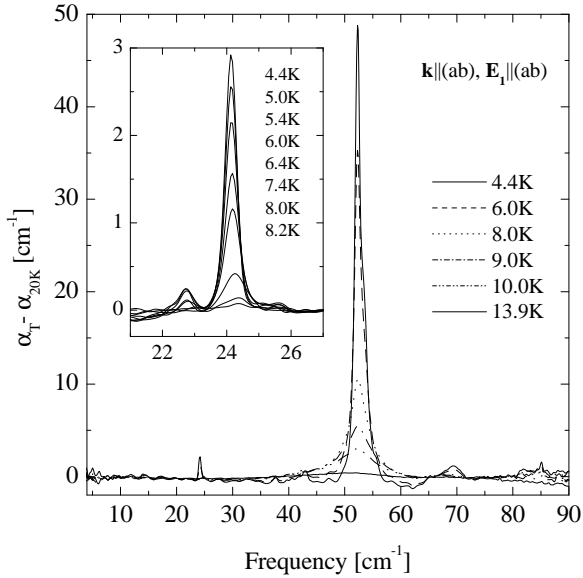, width=7.8cm,
	clip=} }\end{center}

\caption{
Temperature dependence of absorption spectra in  zero magnetic field.
Inset (a) is a  detailed view of the $T$ dependence of the singlet to triplet transition,
$|\mathrm{S}\rangle\rightarrow |\mathrm{T}_0\rangle$.
Side peaks at 22.7 and 25.4\,cm$^{-1}$ are the
$|\mathrm{S}\rangle\rightarrow |\mathrm{T}_{-1}\rangle$
and
$|\mathrm{S}\rangle\rightarrow |\mathrm{T}_{+1}\rangle$  transitions.
}
\end{figure}

Below 10\,K several lines appear between 3 and 100\,cm$^{-1}$ in
the far-infrared transmission spectrum which is rather featureless for the
paramagnetic (20\,K) state (Fig.~1a).
Strong features seen below 30\,cm$^{-1}$ are due to  light interference in
the sample.
Therefore, to see finer details we present in Fig.~1b  spectra
where  the 20\,K spectrum has been subtracted from the 4.4\,K spectrum. 
The transmission does not change by more than  few percents between 20 and 45\,K.
For ($\mathbf{k}\!\parallel\!\mathrm{(ab)}$, $\mathbf{E}_1\!\parallel\!\mathrm{(ab)}$) 
there are   absorption lines  at 24.2, 37.6,  43.0, 52, 69 and 84\,cm$^{-1}$.
The spectrum in ($\mathbf{k}\!\parallel\!\mathrm{c}$, $\mathbf{E}_1\!\parallel\!\mathrm{(ab)}$) 
geometry is similar and is not plotted.
When the electric field is perpendicular to the planes ($\mathbf{k}\!\parallel\!\mathrm{(ab)}$, $\mathbf{E}_1\!\parallel\!\mathrm{c}$)
only one line at 25.4\,cm$^{-1}$ is present.

\begin{figure}
\begin{center}\mbox{ \epsfig{file=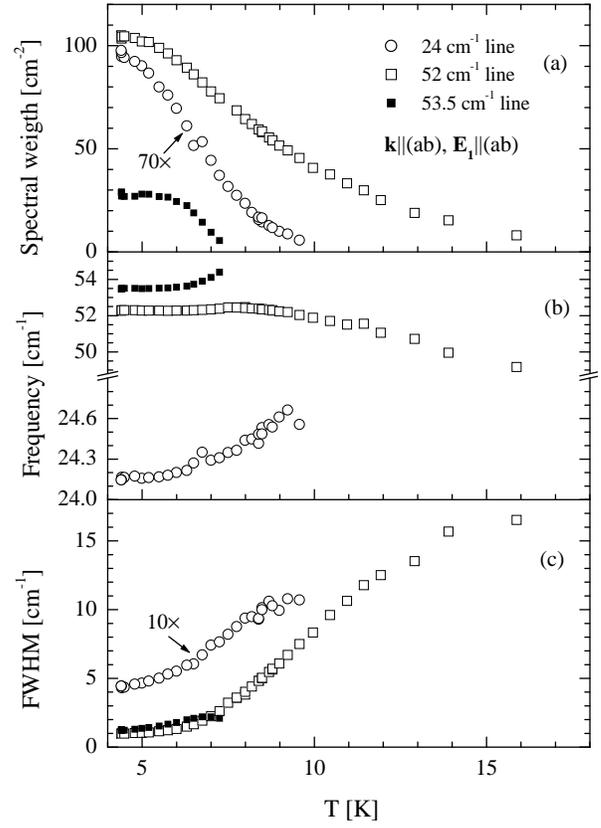, width=7.8cm,
	clip=} }\end{center}

\caption{
Temperature dependencies of the 24, 52 and 53.5\,cm$^{-1}$
lines.
(a) --  the spectral weight (area under the line). Below 8\,K the spectral weight of the 52\,cm$^{-1}$
line is shown as the sum of two Lorentzian lines, one at 52 and the other at 53.5\,cm$^{-1}$. 
(b) --   the line positions.
(c) --  the full width at half maximum (FWHM). 
}
\end{figure}

Temperature dependence of the spectra is shown in Fig.~2
and~3. We have plotted the spectral weight (area under the line),
full width at half maximum (FWHM) and peak positions of the 24.2
and 52\,cm$^{-1}$ lines.
For the 24.2\,cm$^{-1}$ line a better fit was obtained with a Gaussian,
while
a Lorentzian  was used for the 52\,cm$^{-1}$ line.
Analysis of the 52\,cm$^{-1}$ line is complicated by a side peak
at 53.5\,cm$^{-1}$, and  rapid broadening above 8\,K.
Therefore below 8\,K we plot the spectral weight of the 52\,cm$^{-1}$
line as a sum of spectral weights of two Lorentzians located at 52 and 53.3\,cm$^{-1}$.
Above 8\,K we use a single  Lorentzian fit. 

Only the 24.2 and 37.6\,cm$^{-1}$ lines split in  a magnetic field (Fig.4).
The 24.2\,cm$^{-1}$ line corresponds to the transition from the ground
singlet state $|\mathrm{S}\rangle$
 to the  excited triplet state sublevel $|\mathrm{T}_0\rangle$.
This is obvious from the magnetic field dependence of the resonance frequency (Fig~4a).
The line at 25.4\,cm$^{-1}$  seen in Fig.~1b for \textbf{E}$_1$ perpendicular to
the planes
 is the $|\mathrm{S}\rangle\rightarrow |\mathrm{T}_{+1}\rangle$ transition
because its resonace frequency increases with magnetic field (Fig.~4a,4c).
The third component of the triplet resonance, the $|\mathrm{S}\rangle$
 to  $|\mathrm{T}_{-1}\rangle$ transition, is seen when the magnetic field is applied,
\textbf{B}$_0\!\parallel\!\mathrm{(ab)}$, and \textbf{E}$_1\!\parallel\!\mathrm{(ab)}$ (Fig~4c).
There is an additional splitting of the triplet state field dependent
components
(open and filled triangles in Fig.~4a,4c) caused by anisotropic inter-dimer
exchange interaction\cite{Nojiri99}.
Line positions of the  triplet were fitted with
${\mathcal{E}}=\Delta\pm\sqrt{E^2+(g_i\beta B_0)^2}$,
where $\Delta=24.15\pm 0.05$\,cm$^{-1}$, $E=1.4\pm 0.1$\,cm$^{-1}$ and in
plane electron g-factor
$g_a=2.06\pm 0.06$; Bohr magneton $\beta=0.4669$\,cm$^{-1}$T$^{-1}$.
Within the error margins for both geometries
\textbf{E}$_1\!\parallel\!\mathbf{c}$ (Fig.~4a)
and \textbf{E}$_1\!\parallel\!\mathrm{(ab)}$ (Fig.~4c)
the  parameters are the same.

The  resonance line at  37.6\,cm$^{-1}$ splits in the magnetic field  into two components.
From the magnetic field dependence of its resonance frequency we can identify these 
two transitions as transitions from the ground singlet state to the $m_S=-1$  and 
$m_S=+1$  sublevels of the $S\ge1$ multiplet.
The fit  for this multiplet gives slightly
different g-factors  for the upper spin level, $g_u=2.08\pm 0.04$
and  for the lower spin level, $g_l=2.02\pm 0.01$,   with
$\Delta=37.6$\,cm$^{-1}$ and $E=0$.

From the  magnetic field dependence of $S=1/2$ Cu$^{2+}$ spin signal
in the paramagnetic state at 20\,K
we determined the electron g-factor parallel to the planes,
$g_\parallel=2.04\pm 0.04$ and
perpendicular to the planes, $g_\perp=2.33\pm 0.04$.
These are typical values for the
Cu$^{2+}$ spin in a tetragonal crystal field\cite{Abragam70}.
If we normalize the paramagnetic Cu$^{2+}$ signal to the  temperature and energy
splitting of the 24.2\,cm$^{-1}$
singlet-triplet transition in zero magnetic field at 4.4\,K, we get that the 
spectral weight of the singlet to triplet transition 
is  three times
smaller  than that  of the paramagnetic signal.
Magnetic dipole transitions between singlet (antisymmetric) and 
triplet (symmetric) states are  forbidden. 
The antisymmetric Dzialoshinski-Moriya 
interaction\cite{Moriya60} mixes the singlet and triplet
states and therefore makes the singlet-triplet transition possible. 
The local symmetry of a dimer is $C_{2v}$ and consequently 
the Dzialoshinski-Moriya interaction ($J_{DM}$) does not vanish. 
The magnitude of $J_{DM}$ is $J_{DM}\approx(\Delta g/g)J_0$ where 
$\Delta g$ is the deviation of the electron g-factor from the free electron 
value,  $g\approx2$, and $J_0$ is the isotropic exchange interaction. 
If we use $g_c=2.28$ (Ref.\cite{Nojiri99})
and $J_0=24$\,cm$^{-1}$ we get $J_{DM}=2.9$\,cm$^{-1}$.
The zero field splitting of the triplet resonance is 
$2E=2.8$\,cm$^{-1}$, close to the estimate of $J_{DM}$.

Variation of line intensities with magnetic field for the  triplet
state is observed. 
In
$\mathbf{k}\parallel \mathrm{(ab)}, \mathbf{E_1}\parallel \mathbf{c}$
(Fig.~4b) the
$|\mathrm{S}\rangle\rightarrow |\mathrm{T}_0\rangle$
resonance intensity increases from zero to 0.5\,cm$^{-2}$ with increasing magnetic field while
the
$|\mathrm{S}\rangle\rightarrow |\mathrm{T}_{+1}\rangle$
loses its intensity.
The
$|\mathrm{S}\rangle\rightarrow |\mathrm{T}_{-1}\rangle$
resonance intensity   is below noise level.
In
$\mathbf{k}\!\parallel \!\mathrm{(ab)}$, $\mathbf{E_1}\!\parallel
\!\mathrm{(ab)}$ geometry (Fig.~4d)
the 24.2\,cm$^{-1}$ line loses its intensity
whereas transitions to $|\mathrm{T}_{-1}\rangle$  and
$|\mathrm{T}_{+1}\rangle$ levels gain intensity.
The intensity of the 37.6\,cm$^{-1}$ multiplet decreases    nearly linearly with
increasing field (not plotted).
When  the magnetic field is perpendicular to the planes the intensity of
the $|\mathrm{S}\rangle\rightarrow |\mathrm{T}_0\rangle$  transition (24.2\,cm$^{-1}$) does not change with magnetic field. 
Other two components of the singlet to triplet transition, 
 $|\mathrm{S}\rangle\rightarrow |\mathrm{T}_{-1}\rangle$
and $|\mathrm{S}\rangle\rightarrow |\mathrm{T}_{+1}\rangle$, 
 are too weak to be detected in FIR in this field orientation but were observed by ESR\cite{Nojiri99}.
The effect of the magnetic field on the singlet to triplet transition described so far 
is in  changing the oscillator strength of the singlet to triplet transitions.
Magnetic field, besides changing the oscillator strength, has another effect
as shown in Fig.~5. 
Above 8\,T    lines   start to broaden and to lose their intensity.

\begin{figure}
\begin{center}\mbox{ \epsfig{file=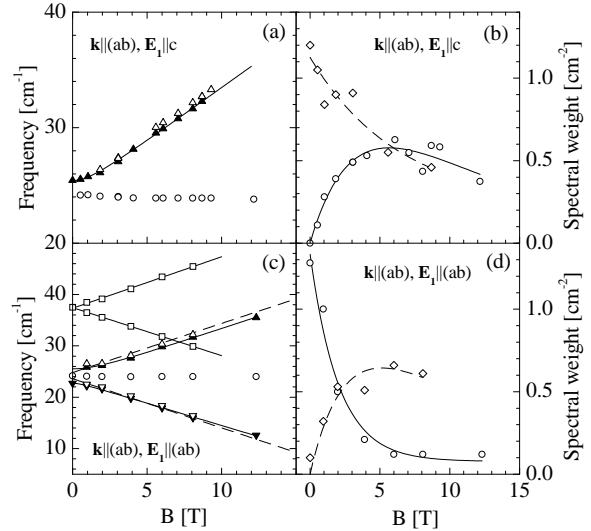, width=7.8cm,
	clip=} }\end{center}

\caption{Magnetic field
$\mathbf{B_0}\!\parallel\!\mathbf{k}$ dependencies of line positions  and
spectral weights
for two light polarizations at 4.4\,K.
Diamonds are the sums of spectral weights of the $|\mathrm{S}\rangle\rightarrow |\mathrm{T}_{-1}\rangle$
and
$|\mathrm{S}\rangle\rightarrow |\mathrm{T}_{+1}\rangle$  transitions
shown by open and filled triangles on (a) and (c) panels.
Solid lines on panels (a) and (c) are the fits described in the text.
Dashed line on panel (c) is the fit of Nojiri \textit{et al.} of their  ESR
data\protect\cite{Nojiri99}.
Lines on panels  (b) and (d) are drawn to guide the eye.}
\end{figure}

The 52\,cm$^{-1}$ resonance is active when the electric field
is parallel to the (ab)-plane independent of the direction of light propagation (Fig.1a). 
Because the light k-vector, $\mathbf{E_1}$  and $\mathbf{H_1}$ fields are orthogonal 
we come to the conclusion that 
the 52\,cm$^{-1}$ resonance is an electric dipole transition. 
This is also supported by the fact that 
the oscillator strength of  the 52\,cm$^{-1}$ resonance 
is considerably larger than the oscillator strength of 
the high temperature paramagnetic Cu$^{2+}$ signal 
or the  24\,cm$^{-1}$ singlet to triplet  transition. 
We exclude the possibility of the 52\,cm$^{-1}$ resonance being a phonon.
The first argument against being a phonon  is the magnetic field effect.
We found that above 8T this line starts to  lose its intensity. 
In the 12T field the spectral weight  is 1.4 times smaller than in  zero field.
The magnetic field effect on this resonance line is very similar to the effect of magnetic field
on the singlet to triplet transition at 24\,cm$^{-1}$ (Fig.~5).
The second argument is the  temperature  dependence.
The  52\,cm$^{-1}$ transition starts to show up below 20\,K (Fig.~3). 
To explain this kind of $T$ dependence for a lattice mode one 
has to assume that a structural phase transition  takes place and the Brillouin zone boundary phonon is folded
back to the zone center. 
However, there is no clear onset temperature of the 52\,cm$^{-1}$ resonance that could be identified 
as a structural phase transition temperature and also the width of the absorption line 
changes by more than one order of magnitude. 
This is not consistent with what is usually observed for  
a zone folded phonon. 
Also, the  Raman scattering experiment\cite{Lemmens99} has not shown any new  phonon
modes below 10\,K and therefore there is no structural phase transition.
We assign the  52\,cm$^{-1}$ line to a transition in the dimer spin system.

Creation of a single quasiparticle in the crystal by  absorption of a light quantum 
is limited to long wavelengths of quasiparticles,
${\mathbf{k}}\approx0$.
Inelastic neutron scattering study\cite{Kageyama2000} has found three branches of magnetic excitations 
in the dimerized state below 20\,K. 
These excitations, in the k-space points that are equivalent to the center of the Brillouin zone,  are at 3, 5 and 9\,meV. 
The 3\,meV excitation corresponds to the singlet to triplet transition  observed in infrared absorption 
at 24\,cm$^{-1}$ and  also by  ESR. 
In the infrared absorption  there are two resonances, 37.6 and 43\,cm$^{-1}$, that are  close in energy 
to the second branch at 5\,meV. Two   higher energy absorption lines at  69\,cm$^{-1}$ and 84\,cm$^{-1}$
 are  close in energy to the third, 9\,meV,  branch. 
No magnetic excitations were detected
in the inelastic neutron scattering experiment 
in the center of the Brillouin zone at 52\,cm$^{-1}$ (6.5\,meV).
We speculate that the 52\,cm$^{-1}$ absorption line in infrared corresponds to an electric dipole active transition 
where two excitations are created simultaneously. 
In this type of transition the spin and the total momentum of two magnetic excitations is zero.
For example,  in antiferromagnets it has been shown 
that the second order electric dipole transition is stronger
than the first order magnetic dipole transition\cite{Loudon68}.

\begin{figure}
\begin{center}\mbox{ \epsfig{file=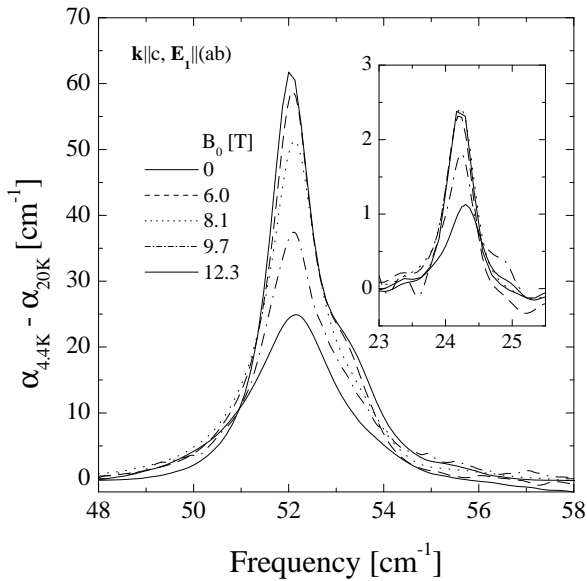, width=7.8cm,
	clip=} }\end{center}
\caption{Magnetic field
 effect on the 24 (inset)  and 52\,cm$^{-1}$
lines at  4.4\,K.}
\end{figure}

The magnetic field suppresses
the singlet to triplet transition and  the 52\,cm$^{-1}$ absorption line (Fig.~5).
In SrCu$_2$(BO$_3$)$_2$ the first gapless ground state is reached 
at 21\,T~(Ref.\cite{Kageyama99,Kageyama98}). 
We observe that already the 9\,T field is affecting line intensities.
Other experiments  have established similar low field effects.
The   magnetization starts to increase from its zero
value in low fields\cite{Kageyama99,Kageyama98}
and in ESR\cite{Nojiri99}  new multiple magnetic resonances have been detected above 12\,T.
Our experiment provides conclusive evidence that 
the magnetic fields smaller than the first critical field change
the gapped ground state of the dimer system.

In conclusion, the spectrum of magnetic excitations  in SrCu$_2$(BO$_3$)$_2$
in the dimerized state has several  infrared active  resonances.
By their magnetic field dependence two of them have been identified as a
triplet resonance at 24\,cm$^{-1}$  and as   a  multiplet at 37.6\,cm$^{-1}$. 
Third is a strong   singlet resonance polarized in the  (ab)-plane at 52\,cm$^{-1}$. 
Other three weak  at 43, 69 and 84\,cm$^{-1}$ are singlet resonances.

We thank professor T.~Timusk and professor H.~Nojiri for helpful
discussions.
Work in Tallinn was partially supported
by Estonian Science Foundation grants no.~3443 and no.~3437.
Work in Japan was supported by a Grant-in-Aid for Encouragement
Young Scientists from The Ministry of Education, Science, Sports and
Culture.



\end{document}